\documentclass[aps,preprint]{revtex4}
\usepackage[dvips]{graphicx}
\usepackage{latexsym}
\usepackage{natbib}

\begin{document}

\title{Interdependent networks with correlated degrees of mutually dependent
nodes}
\author{Sergey V. Buldyrev$^{1}$, Nathaniel Shere$^{1}$, and Gabriel A.
Cwilich$^{1}$}
\affiliation{$^1$Department of Physics, Yeshiva University, 500 West 185th Street, New
York, New York 10033, USA\\
}
\date{\textbf{Revised manuscript: \today }}

\begin{abstract}
We study a problem of failure of two interdependent networks in the case of
correlated degrees of mutually dependent nodes. We assume that both networks
(A and B) have the same number of nodes $N$ connected by the bidirectional
dependency links establishing a one-to-one correspondence between the nodes
of the two networks in a such a way that the mutually dependent nodes have
the same number of connectivity links, i.e. their degrees coincide. This
implies that both networks have the same degree distribution $P(k)$. We call
such networks correspondently coupled networks (CCN). We assume that the
nodes in each network are randomly connected. We define the mutually
connected clusters and the mutual giant component as in earlier works on
randomly coupled interdependent networks and assume that only the nodes
which belong to the mutual giant component remain functional. We assume that
initially a $1-p$ fraction of nodes are randomly removed due to an attack or
failure and find analytically, for an arbitrary $P(k)$, the fraction of
nodes $\mu(p)$ which belong to the mutual giant component. We find that the
system undergoes a percolation transition at certain fraction $p=p_c$ which
is always smaller than the $p_c$ for randomly coupled networks with the same 
$P(k)$. We also find that the system undergoes a first order transition at $%
p_c>0$ if $P(k)$ has a finite second moment. For the case of scale free
networks with $2<\lambda \leq 3$, the transition becomes a second order
transition. Moreover, if $\lambda<3$ we find $p_c=0$ as in percolation of a
single network. For $\lambda=3$ we find an exact analytical expression for $%
p_c>0$. Finally, we find that the robustness of CCN increases with the
broadness of their degree distribution.
\end{abstract}

\maketitle

\section{Introduction}

The robustness of interdependent networks has been recently studied by
Buldyrev \textit{et al.} \cite{Buldyrev} within the framework of the mutual
percolation model. They found that two randomly connected networks with
arbitrary degree distributions randomly coupled by bidirectional dependency
links completely disintegrate via a cascade of failures if the fraction $p$
of the nodes which survive the initial attack is less than some critical
value $p=p_c>0$. Moreover, the transition at $p_c$ is of the first order
type, i.e. the fraction of the functional nodes $\mu(p)$ which survive after
the cascade of failures has a step discontinuity at $p=p_c$ changing from $%
\mu_c=\mu(p_c)>0$ for $p=p_c$ to zero for $p<p_c$. This behavior was
observed even for scale free (SF) networks with a power law degree
distribution $P(k) \sim k^{-\lambda}$ with $2 <\lambda \leq 3$. The
explanation of this behavior is based on the fact that in that model the
nodes with large degree (hubs) in one network may depend on the nodes with
small degree in another network. The nodes with small degree can be isolated
from a giant component in one network by removal of a small fraction of
nodes and thus cause the malfunction of the hubs in the other network. In
real world interacting networks, the hubs in one network are more likely to
depend on the hubs of another networks. This can significantly enhance the
robustness of the interdependent networks. In general, the correlations
among the degrees of the mutually dependent nodes can be described by a
matrix $P(k_1|k_2)$ which specifies the conditional probabilities to find a
node with degree $k_1$ in one network, provided it depends on a node with
degree $k_2$ in another network. This matrix can be quite complex and may
depend on many parameters. For each parameter set the model can be readily
studied by computer simulations, but in order to get general understanding
of the correlation effects it is desirable to solve the problem analytically
at least in some limiting cases.

In this paper we study the mutual percolation problem in the case of the
strongest possible correlations, namely we studied the case in which both
networks (A and B) have the same number of nodes $N$ connected by
bidirectional dependency links establishing a one-to-one correspondence
between the nodes of the two networks in such a way that mutually dependent
nodes have the same number of connectivity links, i.e. their degrees
coincide, i.e. $P(k_1|k_2)=1$ for $k_1=k_2$ and $P(k_1|k_2)=0$ otherwise.
This implies that both networks have the same degree distribution $P(k)$.
For brevity we will call such networks correspondently coupled networks
(CCN), while we will refer to the model studied in Ref.~\cite{Buldyrev} as
randomly coupled networks (RCN). As in Ref.~\cite{Buldyrev}, we assume that
the nodes in each network are randomly connected. We define the mutually
connected clusters and the mutual giant component as in Ref.~\cite{Buldyrev}
and assume that only the nodes which belong to the mutual giant component
remain functional.

We assume that initially a $1-p$ fraction of nodes are randomly removed due
to an attack or failure and find analytically the fraction of nodes $\mu(p)$
which belong to the mutual giant component. We find that as in Ref.~\cite%
{Buldyrev}, the system undergoes a percolation transition at certain
fraction $p=p_c$ which, however, is always smaller than the $p_c$ for RCN
with the same degree distribution with the exception of random regular
graphs~\cite{bollo} for which both values coincide. Moreover, we find that
the system undergoes a first order transition at $p_c>0$ if the degree
distribution has a finite second moment. For the practically imqportant case
of SF networks\cite{barasci,bararev,PastorXX,mendes,cohena} with $2<\lambda
\leq 3$, for which the second moment diverges, the transition becomes a
second-order transition. If $\lambda<3$ we find that $p_c=0$ as in
percolation of a single network\cite{cohen2000}, while for $\lambda=3$ we
find an exact analytical expression for $p_c>0$. The  change in transition order has been also observed
in interdependent networks with partial coupling \cite{Parshani}. We also investigate how the
broadness of the degree distribution affects the robustness of CCN.

\section{Generating functions and the cascade process}

\subsection{First stage}

We will describe the stages of the cascade of failures in CCN in terms of
the generating function of their degree distribution~\cite{Newman,mnewman} 
\begin{equation}  \label{G}
G(x) = \sum_{k=0}^{\infty} P(k)x^k
\end{equation}
and the generating function of the associated branching process~\cite%
{Harris1} 
\begin{equation}  \label{H}
H(x)=\frac{G^{\prime }(x)}{G^{\prime }(1)}= \frac{1}{\langle k \rangle}\frac{%
dG(x)}{dx},
\end{equation}
where $\langle k\rangle \equiv G^{\prime }(1)$ is the average degree. It is
known that the degree distribution $\tilde{P}(k,p)$ of a network, from which
a fraction of nodes $1-p$ is randomly removed is related to the original
distribution $P(k)$ through a binomial expansion~\cite{Newman}: 
\begin{equation}
\tilde{P}(k^{\prime },p)=\sum_{k\geq k^{\prime }}P(k)p^{k^{\prime
}}(1-p)^{k-k^{\prime }}C_k^{k^{\prime }},  \label{tildeP}
\end{equation}
where $C_k^{k^{\prime }}=k!/[k^{\prime }!(k-k^{\prime })!]$ are binomial
coefficients. Accordingly~\cite{Newman}, the generating function of this
distribution is 
\begin{equation}  \label{tildeG}
\tilde{G}(x,p)=G(xp +1-p).
\end{equation}
The fraction of nodes which do not belong to the giant component of a
network is given by~\cite{Shao,Shao08} 
\begin{equation}  \label{r}
r=G(f),
\end{equation}
where $f$ is a smallest nonnegative root a transcendental equation 
\begin{equation}  \label{f}
f=H(f).
\end{equation}
The degree distribution of nodes which do not belong to the giant component
is given by \cite{Shao} 
\begin{equation}  \label{Po}
P_o(k,f)=P(k)f^k/r.
\end{equation}
Accordingly the degree distribution of nodes in the giant component is given
by 
\begin{equation}  \label{Pi}
P_i(k,f)=P(k)(1-f^k)/(1-r).
\end{equation}
Thus the degree distribution in the giant component of a decimated network
after random removal of a $1-p$ fraction of nodes is 
\begin{equation}  \label{tildePi}
\tilde{P}_i(k^{\prime },f,p)=\tilde{P}(k^{\prime},p)[1-f(p)^{k^{\prime }}]/[1-r(p)].
\end{equation}
where 
\begin{equation}  \label{rp}
r(p)=\tilde{G}(f(p),p)
\end{equation}
and $f(p)$ satisfies the transcendental equation 
\begin{equation}  \label{fp}
f(p)=\tilde{H}(f(p),p).
\end{equation}
In order to find the original degree distribution in the giant component of
network A we must restore the links which lead to the randomly removed
nodes. If a node in the decimated network A has a degree $k^{\prime }$ it
might have any degree $k\geq k^{\prime }$ in the original network A with
probability $P(k|k^{\prime })$ given by the Bayes' formula: 
\begin{equation}  \label{bayes}
P(k|k^{\prime })=P(k)C_k^{k^{\prime }}p^{k^{\prime }}(1-p)^{k-k^{\prime }}/%
\tilde{P}(k,p).
\end{equation}
Thus the total probability that a node in the giant component has a degree $k
$ is 
\begin{equation}  \label{total1}
P_1(k)=\sum_{k^{\prime }\leq k}P(k)C_k^{k^{\prime }}p^{k^{\prime
}}(1-p)^{k-k^{\prime }}\frac{\tilde{P}_i(k^{\prime },f,p) }{\tilde{P}%
(k^{\prime },p)},
\end{equation}
or using Eq. (\ref{tildePi}) 
\begin{equation}  \label{total}
P_1(k)=\sum_{k^{\prime }\leq k}P(k)C_k^{k^{\prime }}p^{k^{\prime
}}(1-p)^{k-k^{\prime }}\frac{1-f(p)^{k^{\prime }}}{1-r(p)}=P(k)\frac{%
1-(f(p)p+1-p)^k}{1-r(p)}.
\end{equation}
The generating function of this degree distribution is 
\begin{equation}  \label{G1}
G_1(x)=\frac{G(x)-G(xt_1)}{1-G(t_1)},
\end{equation}
where $t_1=f_1p+1-p$ and $f_1=f(p)$. The fraction of nodes in the giant
component of the decimated network A is $1-r_1$, where $r_1=G(t_1)$, Because
the decimated network has $Np$ nodes, the size of the giant component A$_1$
of network A after random removal of $(1-p)$ nodes is $N_1=Np(1-r_1)$.

\subsection{Second stage}

We assume that only nodes which belong to A$_1$ are functional; thus after
the first stage of the cascades of failures, only $p(1-r_1)<p$ fraction of
nodes in network B remain functional. Thus we expect further disintegration
of network B at the second stage of the cascade and its giant component B$_2$
will be even smaller than A$_1$. We define a set of nodes B$_1 =D($A$_1)$ by
projecting A$_1$ onto network B using the one-to-one correspondence $D$
between the nodes of networks A and B established by dependency links. Since
the degree of each node in network B is the same as the degree of its
dependent node in network A, the giant component A$_1$ of network A obtained
at the first stage of the cascade has the same degree distribution as the
set B$_1$ in network B. Moreover, from the point of view of network B the
nodes in B$_1$ are randomly selected and randomly connected. Thus, to
compute B$_2$ we can use the same approach used at the first stage, but
applied to the new network B$_1$ with the new degree distribution given by
Eq. (\ref{total}). The only problem is that many of the links outgoing from
network B$_1$ are ending at the nodes which do not belong to network B$_1$
and thus for computation of B$_2$ these links must be removed. The
probability $p_1$ of a random link originating in network B$_1$ to end up in
B$_1$ is equal to the ratio of the number of links originating in network B$%
_1$: 
\begin{equation}
L_1=N_1\sum k P_1(k)=p N\langle k \rangle(1- G^{\prime }(t_1)t_1/\langle
k\rangle)
\end{equation}
to the total number of links $N\langle k \rangle$. Therefore, 
\begin{equation}  \label{p1}
p_1=\frac{L_1}{N\langle k\rangle}=p(1-s_1),
\end{equation}
where 
\begin{equation}  \label{s1}
s_1=t_1G^{\prime }(t_1)/\langle k \rangle .
\end{equation}
Accordingly, the degree distribution of links connecting the nodes of
network B$_1$ is 
\begin{equation}  \label{tildeP1}
\tilde{P}_1(k^{\prime },p)=\sum_{k\geq k^{\prime }}P_1(k)p_1^{k^{\prime
}}(1-p_1)^{k-k^{\prime }}C_k^{k^{\prime }}
\end{equation}
and the generating function of this distribution is 
\begin{equation}  \label{tildeG1}
\tilde{G}_1(x,p_1)=[G(xp_1 +1-p_1)-G(t_1(xp_1+1-p_1))]/(1-r_1).
\end{equation}
Thus the size $N_2$ of the giant component B$_2$ is $N_2=p(1-r_1)(1-r_2)N$,
where $r_2=\tilde{G}_1(f_2,p_1)$ and $f_2=\tilde{H}_1(f_2,p_1)$. Introducing
a new notation 
\begin{equation}  \label{t2}
t_2 \equiv f_2p_1+1-p_1
\end{equation}
and taking into account Eq. (\ref{tildeG1}) we see that 
\begin{equation}  \label{f2}
f_2=\frac{G^{\prime }(t_1)-G^{\prime }(t_1t_2)t_1}{\langle k\rangle (1-s_1)}
\end{equation}
and $N_2=p(1-r_1)[1-(G(t_2)-G(t_2t_1))/(1-r_1)]$. Using that $r_1=G(t_1)$,
we get 
\begin{equation}  \label{N2}
N_2= p[1-G(t_1)-G(t_2)+G(t_1t_2)]N
\end{equation}
We can compute the original degree distribution $P_2(k)$ in B$_2$ using the
Bayes' formula the same way as we obtained the distribution $P_1(k)$ 
\begin{equation}  \label{total2}
P_2(k)=P(k)\frac{(1-t_1^k)(1-t_2^k)}{1-G(t_1)-G(t_2)+G(t_1t_2)}.
\end{equation}

\subsection{Third stage}

On the third stage of the cascade we will compute the giant component A$_{3}$
of network A which is the result of further disintegration of A$_{1}$
because the nodes in A$_{1}$ which do not belong to B$_{2}$ failed at the
second stage. We can again apply the same technique, with the only
difference that now A$_{2}=D($B$_{2})$ is not a random subset of nodes of A
but they are taken out of its connected giant component A$_{1}$.
Accordingly, we must find an effective degree distribution and the effective
size of a network which would reproduce A$_{2}$ by random selection of nodes
out of this original network. Since the degree distribution of nodes in A$%
_{1}$ is 
\begin{equation}
P_{1}(k)=P(k)\frac{1-t_{1}^{k}}{1-G(r_{1})}  \label{total1a}
\end{equation}%
and the degree distribution in A$_{2}$ is 
\begin{equation}
P_{2}(k)=P(k)\frac{(1-t_{2}^{k})(1-t_{1}^{k})}{(1-r_{2})(1-r_{1})},
\label{total2a}
\end{equation}%
the selection of A$_{2}$ out of A$_{1}$ has the same effect as random
selection of a fraction $p(1-r_{1})$ out of the entire network with degree
distribution $P_{2}^{\ast }(k)=P(k)(1-t_{2}^{k})/(1-r_{2})$. Now we can see
that the problem of the third stage is completely equivalent to the second
stage with $t_{1}$ replaced by $t_{2}$, $f_{1}$ replaced by $f_{2}$, $r_{1}$
replaced by $r_{2}$, $t_{2}$ replaced by $t_{3}$, $f_{2}$ replaced by $f_{3}$%
, and $r_{2}$ replaced by $r_{3}$.

\subsection{Recursive relations}

Generalizing, for stage $i$ we arrive to a recursive relation between $t_{i}$
and $t_{i+1}$. Namely, once we know $t_{i}$ we can find $t_{i+1}$, as well
as the size of the giant component at the stage $i+1$ 
\begin{equation}
N_{i+1}=p[1-G(t_{i})-G(t_{i+1})+G(t_{i}t_{i+1})]N  \label{Ni}
\end{equation}%
and the degree distribution of the nodes inside this giant component 
\begin{equation}
P_{i+1}(k)=P(k)\frac{(1-t_{i+1}^{k})(1-t_{i}^{k})}{%
1-G(t_{i})-G(t_{i+1})+G(t_{i}t_{i+1})}  \label{totali}
\end{equation}%
In order to find $t_{i+1}$ from $t_{i}$ we repeat the steps used deriving $%
t_{2}$ from $t_{1}$ by first introducing 
\begin{equation}
s_{i}=t_{i}G^{\prime }(t_{i})/\langle k\rangle ,  \label{si}
\end{equation}%
and 
\begin{equation}
p_{i}=p(1-s_{i})  \label{pi}
\end{equation}%
in analogy to Eqs. (\ref{si}) and (\ref{p1}). Then 
\begin{equation}
t_{i+1}\equiv f_{i+1}p_{i}+1-p_{i},  \label{ti}
\end{equation}%
where $f_{i+1}$ satisfies a transcendental equation analogous to Eq.~(\ref%
{f2}) 
\begin{equation}
f_{i+1}=\frac{G^{\prime }(t_{i})-G^{\prime }(t_{i}t_{i+1})t_{i}}{\langle
k\rangle (1-s_{i})}.  \label{fi}
\end{equation}%
Excluding $f_{i+1}$ and $s_{i}$ from Eq.~(\ref{fi}) we find that $t_{i+1}$
is given by the smallest non-negative root of the equation 
\begin{equation}
t_{i+1}=(1-p)+\frac{p}{\langle k\rangle }[G^{\prime }(t_{i})t_{i}+G^{\prime
}(t_{i+1})-t_{i}G^{\prime }(t_{i}t_{i+1})].  \label{ti1}
\end{equation}%
To start the iterative process we must take into account the definition of $%
t_{1}=pf_{1}+1-p$ and Eq. (\ref{fp}) which is equivalent to a transcendental
equation 
\begin{equation}
t_{1}=(1-p)+\frac{p}{\langle k\rangle }G^{\prime }(t_{1}),  \label{t1}
\end{equation}%
which is the same as Eq. (\ref{ti1}) if we introduce $t_{0}\equiv 0$.

\section{The mutual giant component and the phase transition}

The cascade of failures will stop when $t_{i+1}=t_{i}=t$ and hence the
fraction of nodes in the mutual giant component $\mu=\lim_{i\to \infty}N_i/N$%
, is given by the simplified equation (\ref{Ni}): 
\begin{equation}  \label{mu}
\mu= p[1-2G(t)+G(t^2)],
\end{equation}
where $t$ is the smallest non-negative root of the equation 
\begin{equation}  \label{t}
t =(1-p)+ \frac{p}{\langle k \rangle }[(1+t)G^{\prime }(t)-tG^{\prime
}(t^2)]=1-p[1-(1+t)H(t)+tH(t^2)].
\end{equation}

The right hand side of Eq. (\ref{t}) has zero derivative at $t=1$, if $%
G^{\prime \prime }(1)$ is finite. This condition is equivalent to the
existence of the second moment of the degree distribution. Thus one can see
[Fig.~\ref{fig1}(a,b)] that for finite second moment and small enough $p$,
Eq. (\ref{t}) has only the trivial solution $t=1$ corresponding to $\mu=0$
and, therefore, to the complete disintegration of the networks. As $p$
increases, a nontrivial solution $\mu>0$ will emerge at $p=p_c$, at which
point the right hand side of Eq. (\ref{t}) will touch the straight line
representing the left hand side at $t=t_c$; at that point the slope of both
lines is equal to 1. Since at $t=1$ the slope of the right hand side is
zero, $t_c$ must be smaller than 1 and thus the mutual percolation
transition is of the first order, where $\mu$ changes form zero (for $p<p_c$%
) to $\mu\geq \mu_c>0$ (for $p \geq p_c$). The value of $\mu_c$ is given by
Eq.~(\ref{mu}) computed at $t=t_c$.

An efficient way of finding $p_c$ is to solve Eq. (\ref{t}) with respect to $%
1/p$: 
\begin{equation}  \label{1p}
\frac{1-(1+t)H(t)+tH(t^2)}{1-t}=\frac{1}{p}
\end{equation}
and find the maximum of the left hand side with respect to $t$ [Fig.~\ref%
{fig2}]. The left hand side of Eq. (\ref{1p}) is a curve which changes from $%
1-H(0)=1-P(1)/\langle k \rangle$ at $t=0$ to zero at $t=1$. At $t=0$ it has
a positive slope $1-(P(1)+2P(2))/\langle k \rangle)$, so it must have an
absolute maximum at $t_c\in (0,1)$. The equation for $t_c$ can be readily
obtained by differentiation of Eq. (\ref{1p}) 
\begin{equation}  \label{tc}
1-2H(t_c)+H(t_c^2)-(1-t_c^2)H^{\prime }(t_c)+2t_c^2(1-t+c)H^{\prime
}(t_c^2)=0.
\end{equation}

The value of the left hand side of Eq. (\ref{1p}) at $t=t_c$ gives $1/p_c$.
If the value of this maximum is less than 1, than the networks do not have a
mutual giant component at any $p$.

\section{Special Cases}

Figure \ref{fig1} shows the graphical solutions of Eq. (\ref{t}) for several
special cases of degree distributions of CCN.

\subsection{Erd\H{o}s-R\'{e}nyi networks}

For Erd\H{o}s-R\'{e}nyi (ER) networks\cite{bollo,er1} $H(t)=\exp[\langle
k\rangle(t-1)]$ and the maximal value of the left hand side of Eq.~(\ref{1p}%
) monotonously increases with $\langle k \rangle$. This can be readily seen
by differentiating Eq.~(\ref{1p}) with respect to $\langle k \rangle$. The
maximal value reaches 1 at $\langle k \rangle=1.706526$, below which
correspondently coupled ER networks disintegrate even without any initial
attack or failure (Fig.~\ref{fig2}). Note that the equivalent value of $%
\langle k \rangle$ for randomly coupled ER networks is 2.4554\cite{Buldyrev}.

\subsection{Random regular graphs}

For RR graph [Fig. \ref{fig1}(b)] in which all the nodes have the same
degree $k=\langle k\rangle $, $G(t)=t^{\left\langle k\right\rangle }$ and $%
H(t)=t^{\langle k\rangle -1}$. Then $t$ satisfies 
\begin{equation}
t=(1-p)+p(t^{\langle k\rangle -1}+t^{\langle k\rangle }-t^{2\langle k\rangle
-1}),  \label{trr}
\end{equation}%
and 
\begin{equation}
\mu =p(1-t^{\left\langle k\right\rangle })^{2}.  \label{murr}
\end{equation}%
Equations (\ref{trr}) and (\ref{murr}) can be obtained by simpler methods
presented in Ref. \cite{Buldyrev} for RCN, since for the case of random
regular graphs, the degrees of all the nodes in both networks coincide and
therefore the CCN and RCN models are equivalent. Indeed from Eq. (1) of Ref. 
\cite{Buldyrev} it follows in a special case of coinciding degree
distributions of the coupled networks that 
\begin{equation}
\mu =p(1-G(t))^{2},  \label{mub}
\end{equation}%
where 
\begin{equation}
t=1-p(1-G(t))(1-H(t)).  \label{tb}
\end{equation}%
If $G(t)=t^{\left\langle k\right\rangle }$ and $H(t)=t^{\langle k\rangle -1}$%
, Eqs. (\ref{mub}) and (\ref{tb}) are equivalent to Eqs. (\ref{murr}) and (%
\ref{trr}), respectively.

\subsection{ Scale free networks}

\begin{figure*}[h]
\centerline{
    \includegraphics[width=4.5cm,angle=0]{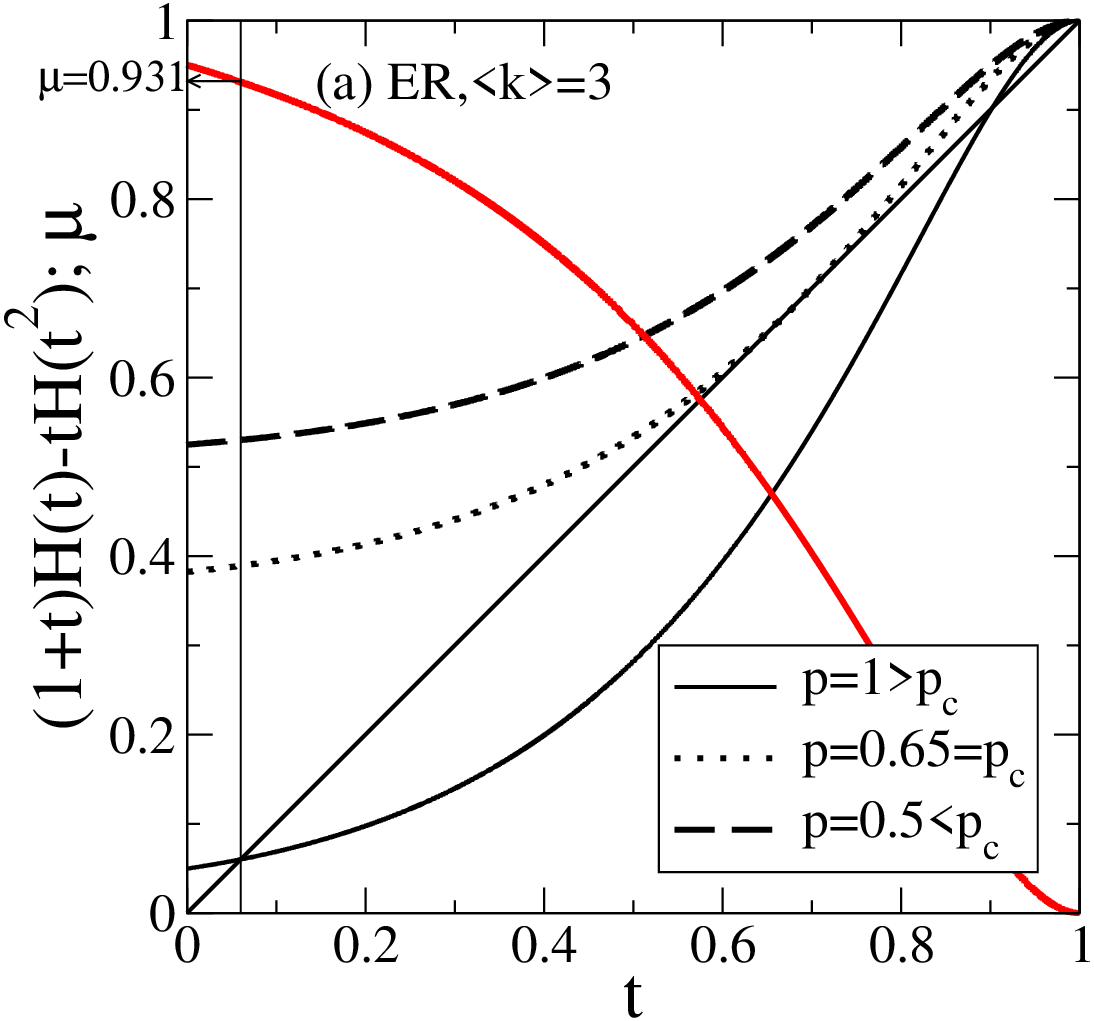}
 \includegraphics[width=4.5cm,angle=0]{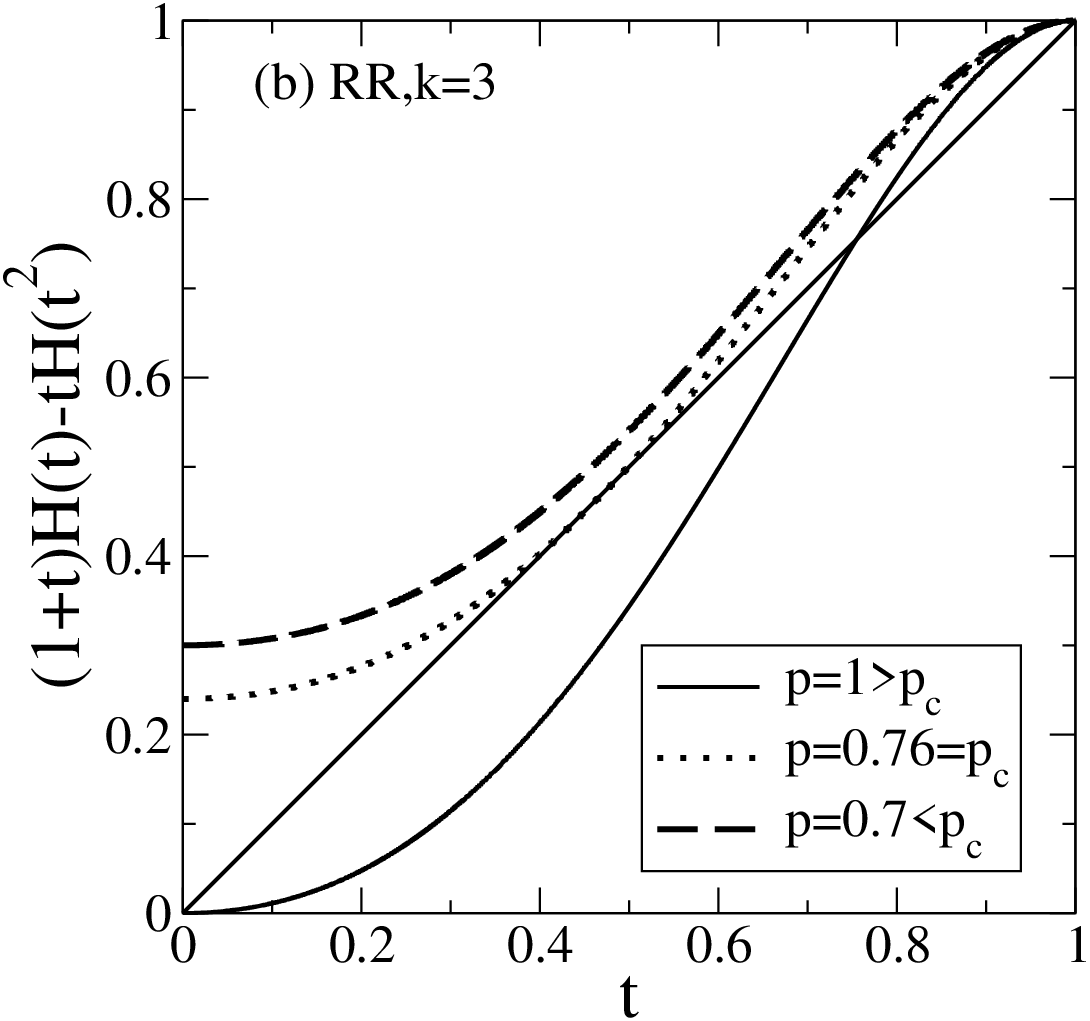}} 
\centerline{
 \includegraphics[width=4.5cm,angle=0]{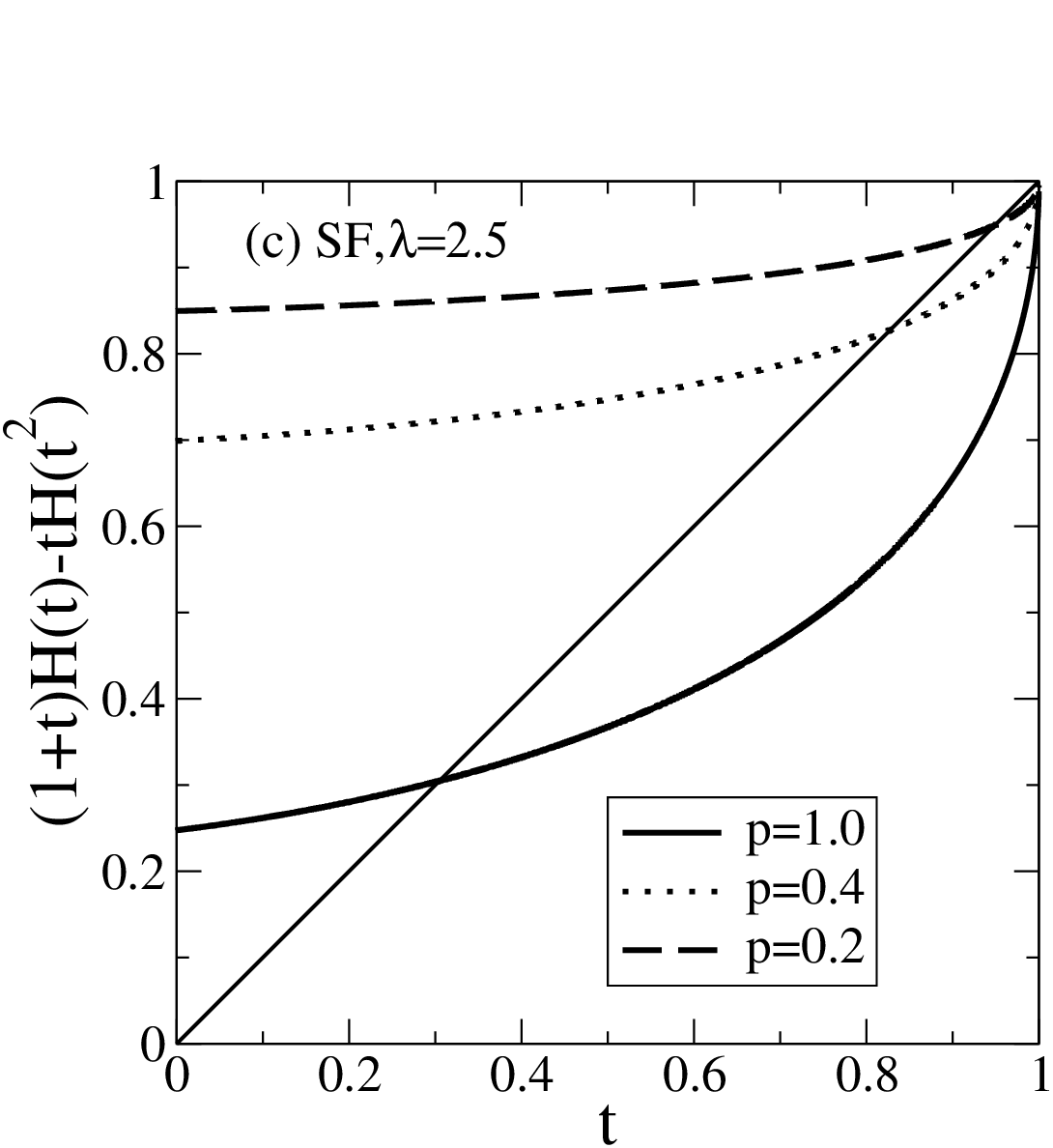}
 \includegraphics[width=4.5cm,angle=0]{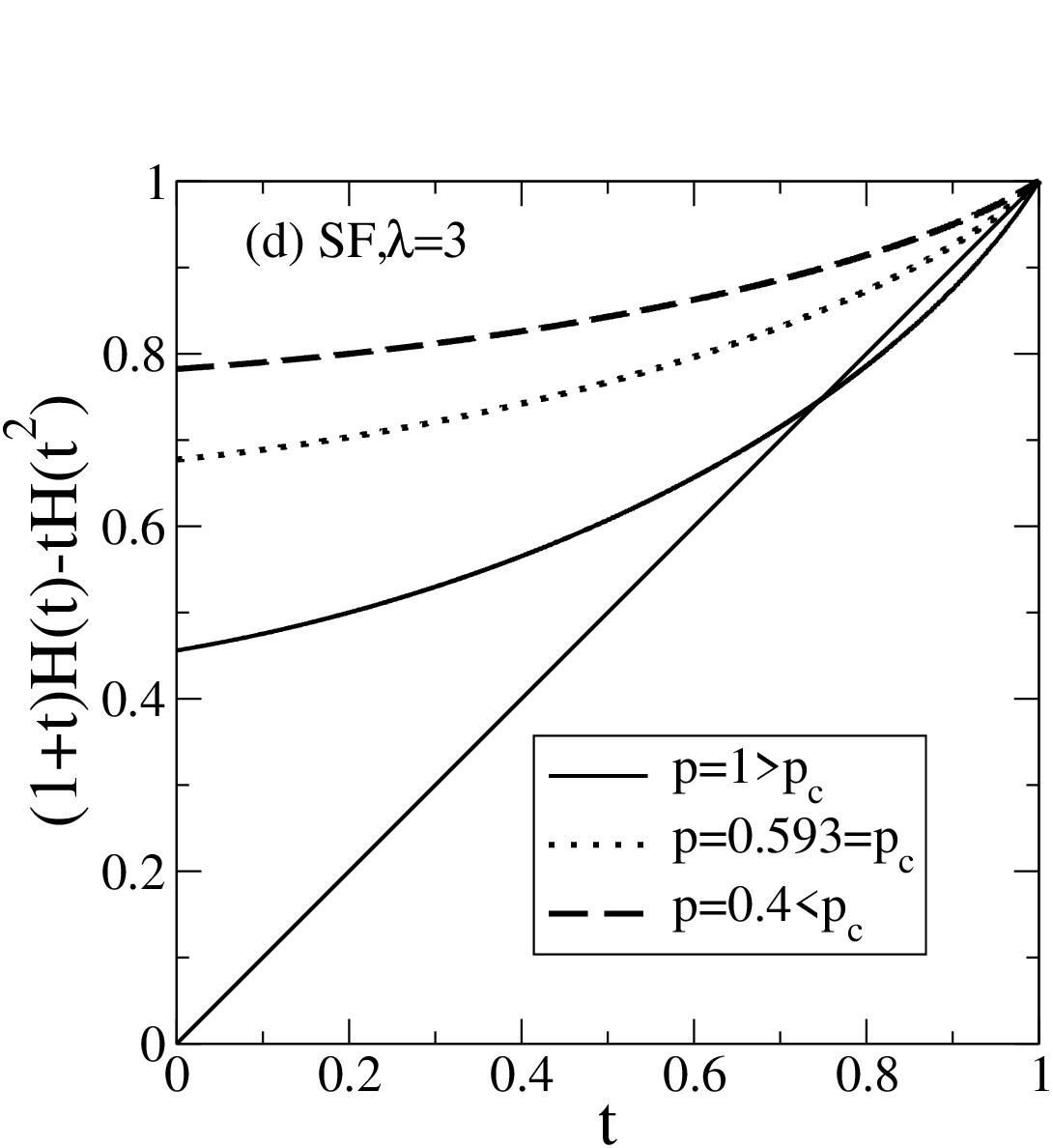}} 
\caption{ Graphical solution of Eq. (\protect\ref{t}) for various special
cases of CCN. (a) ER networks with average degree $\langle k\rangle =3$. One
can see that the black curves representing the right hand side of Eq. (%
\protect\ref{t}) for different $p$ have zero slopes at $t=1$. The relevant
solutions for $t$ are given by the lower intersection points of these curves
and a straight line $y=t$ representing the left hand side of Eq.~(\protect
\ref{t}). For $p=1$, this solution $t=0.0602$ is indicated by a vertical
straight line. The intersection of this vertical line with the plot of Eq. (%
\protect\ref{mu}) (red curve) gives the mutual giant component $\protect\mu %
=0.931$. The critical $p=p_{c}=0.6499451$ corresponds to sudden
disappearance of the nontrivial solution. (b) RR networks with $k=3$. Note
that for $p=1$ the nontrivial solution is $t=0$ which means that $\protect%
\mu =1$. The value of $p_{c}=0.758751$ is grater than the $p_{c}$ for ER
networks with the same average degree shown in panel (a). (c) Analogous plot
for scale free networks with $\protect\lambda =2.5$. It shows that the slope
of the curves is infinite for $t\rightarrow 1$. One can see that in this
case the nontrivial solution exists for any $p>0$. However as $p\rightarrow 0
$, the nontrivial solution $t\rightarrow 1$, and, accordingly, $\protect\mu %
\rightarrow 0$ indicating the second order transition at $p=p_{c}=0$. (d)
The marginal case of $\protect\lambda =3$. The slopes of the curves for $%
t\rightarrow 1$ are finite. This means that there is a critical $p=p_{c}>0$
at which the slope of the curve becomes equal to 1 at $t\rightarrow 1$. For
the displayed case of $k_{min}=1$, Eq. (\protect\ref{pc2}) yields $%
p_{c}=0.59328456$. The nontrivial solution smoothly approaches 1 as $%
p\rightarrow p_{c}$. This again implies $\protect\mu \rightarrow 0$ (second
order transition). }
\label{fig1}
\end{figure*}

\begin{figure*}[h!]
\includegraphics[width=8cm,angle=0]{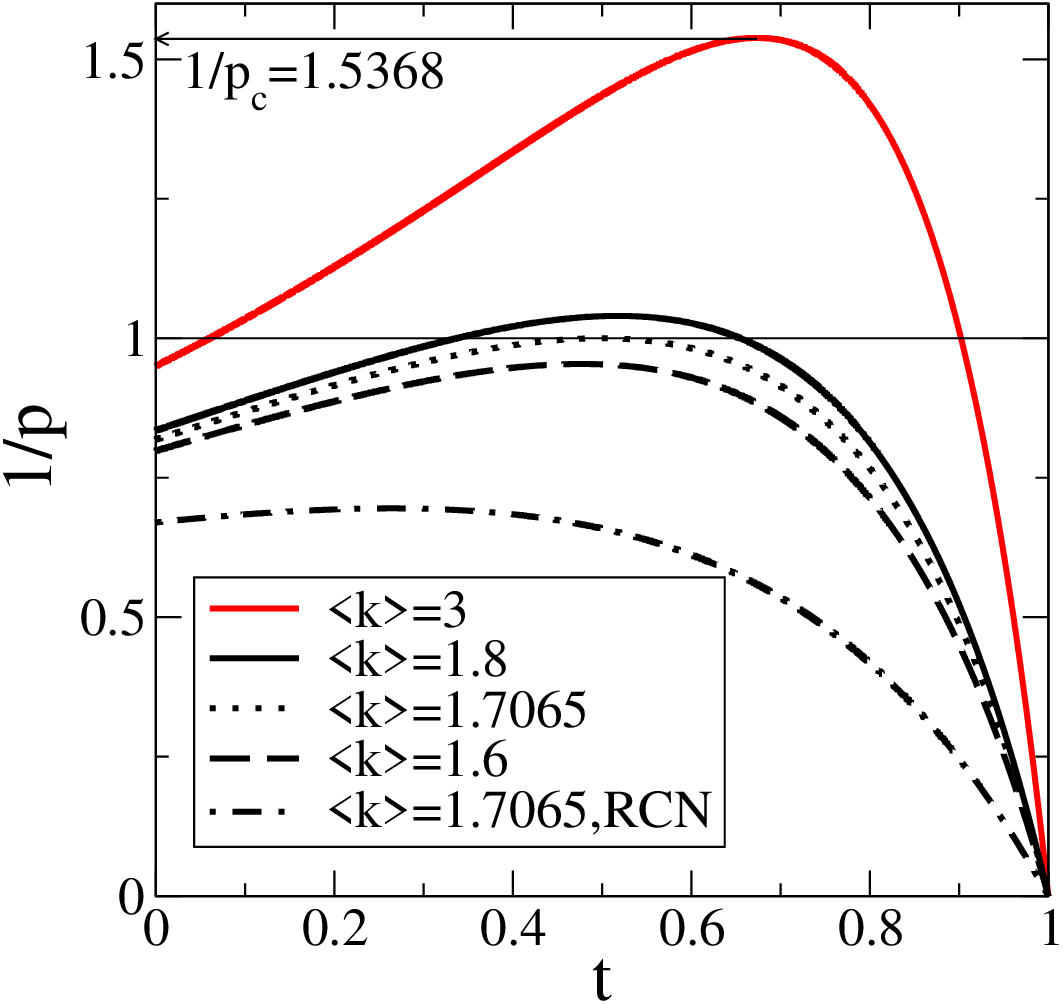} 
\caption{ Graphical solution of Eq. (\protect\ref{1p}) for ER networks with
different degree $<k>$ illustrating the method of finding $p_c$. The red
curve corresponds to $\langle k\rangle=3$ studied in Fig.\protect\ref{fig1}%
(a). As $<k>$ decreases below 1.706, the nontrivial solution corresponding
to $p \leq 1$ disappears. We also show the behavior of the analogous
equation (\protect\ref{1pb}) for $\langle k\rangle =1.706$ for RCN. In
agreement with proposition (\textit{iii}) this curve is always below the
curve with the same average degree for networks with coinciding degrees
studied here.}
\label{fig2}
\end{figure*}

For scale free (SF) networks with $\lambda<3$ [Fig.~\ref{fig1}(c)], the
second derivative of the right hand side of Eq. (\ref{t}) is infinite at $t=1
$, which means that a nontrivial solution exists at any $p>0$ since in the
vicinity of $t=1$ the straight line representing the left hand side of Eq. (%
\ref{t}) is always above the curve representing the right hand side, while
for $t=0$, the curve is always above the line. This means that SF CCN are as
robust as a single SF network for which $p_c$ is always zero.

For the marginal case of $\lambda =3$ [Fig.~\ref{fig1}(d)] $G^{\prime \prime
}(t)$ diverges as $\ln (1-t)$ when $t\rightarrow 1$ and thus the left hand
side of Eq. (\ref{t}) has a finite slope at $t=1$. Accordingly $p=p_{c}>0$
but the nontrivial solution emerges at $t_{c}=1$, so the transition becomes
of the second order. For the case of $%
P(k)=(k_{min}/k)^{2}-(k_{min}/(k+1))^{2}$ for $k\geq k_{min}=1,2,..$ and $%
P(k)=0$ for $k<k_{min}$ we can find $p_{c}$ analytically. Indeed, in this
case $P(k)$ behaves asymptotically as $2k_{min}^{2}/k^{3}$. For $%
k\rightarrow \infty $ the leading term in $G^{\prime \prime }(t)$ becomes $%
2k_{min}^{2}t^{k}/k$; so,  $G^{\prime \prime }(t)=-2k_{min}^{2}\ln (1-t)+c(t)
$, where $c(t)$ is a continuous function. Accordingly, the slope of the
right hand side of Eq.~(\ref{t}) at $t=1$ becomes $p4k_{min}^{2}\ln
(2)/\langle k\rangle $, where 
\begin{equation}
\langle k\rangle =k_{min}+k_{min}^{2}\left( \frac{\pi ^{2}}{6}%
-\sum_{k=1}^{k_{min}}\frac{1}{k^{2}}\right) .  \label{avk}
\end{equation}%
The critical threshold is thus 
\begin{equation}
p_{c}=\frac{\frac{1}{k_{min}}+\frac{\pi ^{2}}{6}-\sum_{k=1}^{k_{min}}\frac{1%
}{k^{2}}}{4\ln (2)}  \label{pc2}
\end{equation}%
For $k_{min}=1$ $p_{c}=0.59328456$ and for $k_{min}=2$ $p_{c}=0.32277924$.

\subsection{Effect of the broadness of the degree distribution}

It follows from Fig.~\ref{fig1} that for the same $\langle k\rangle =3$, $%
p_{c}$ of the RR networks (0.758751) is greater than the $p_{c}$ of the ER
networks (0.6499451). Moreover, for SF networks with $\lambda =3$ and $%
k_{min}=1$, for which the average degree is $\pi ^{2}/6<3$ we have even
smaller $p_{c}=0.59328456$. For SF networks with $\lambda =3$, $\langle
k\rangle =3$ we can estimate $p_{c}=0.35$ which is much smaller than the $%
p_{c}$ for the narrower ER and RR degree distribution. For SF networks with $%
\lambda <3$, which are even broader, $p_{c}=0$ for any $\langle k\rangle $.
This is in a complete agreement with the trend observed in percolation of
single networks, for which the robustness increases with the broadness of
the degree distribution if one keeps $\langle k\rangle $ constant, but is
opposite to the trend observed in Ref. \cite{Buldyrev} for RCN.

\begin{figure*}[h!]
\includegraphics[width=8cm,angle=0]{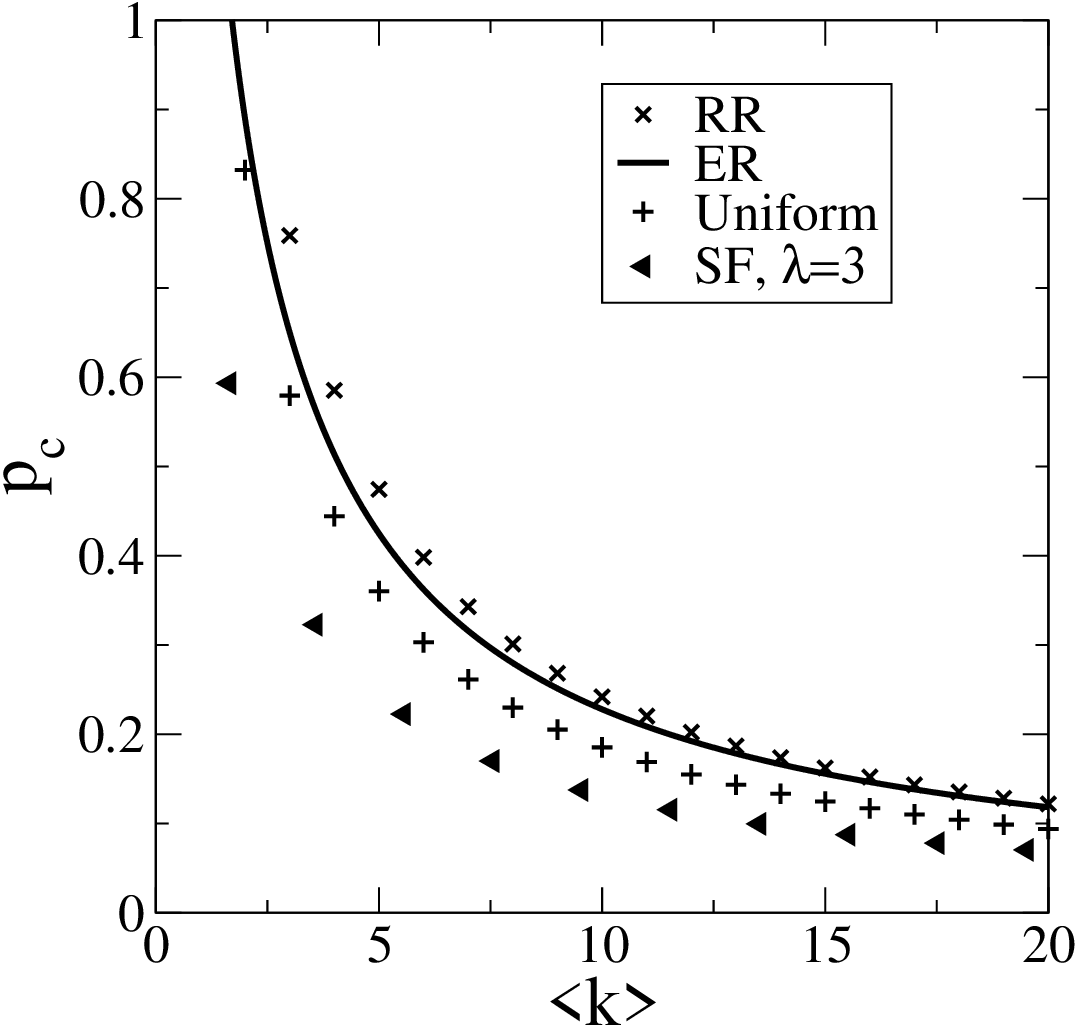} 
\caption{ The values of $p_c$ versus $\langle k\rangle$ for several degree
distributions of increasing broadness, namely RR, ER, uniform and SF with $%
\protect\lambda=3$. We define the uniform distribution as follows: $%
P(k)=1/(2\langle k\rangle+1)$ for $k=0,1,...,2\langle k\rangle$ and $P(k)=0$
for $k>2\langle k\rangle$. For SF distribution we use Eqs. (\protect\ref{avk}%
) and (\protect\ref{pc2}), while for other distributions we numerically
solve Eq. (\protect\ref{tc}) and use Eq. (\protect\ref{1p}) to find $p_c$.
One can see that $p_c$ decreases ( and hence the robustness increases) with
the broadness.}
\label{fig3}
\end{figure*}

In order to investigate the effect thoroughly, we study several classes of
degree distributions for a number of values of $\langle k \rangle$. Figure %
\ref{fig3} shows $p_c$ as function of $\langle k \rangle$ for RR, ER,
uniform, and SF with $\lambda=3$ degree distributions. For each value of $%
\langle k \rangle$ the variance of SF degree distribution ($\infty$) is
greater than the variance of the uniform degree distribution ($\langle k
\rangle(\langle k\rangle +1)/3$), which is greater than the variance of ER
degree distribution $\langle k \rangle$ which is greater than the variance
of RR degree distribution (0). Indeed, Fig.~\ref{fig3} shows that $p_c($SF$%
)<p_c($uniform$)<p_c($ER$)<p_c($RR$)$. Thus our numerical results suggest
that CCN become more robust if their degree distribution becomes broader
(provided the average degree is constant). This behavior is the opposite of
the behavior of RCN.

However, in general, if the measure of broadness is simply the variance of
the degree distribution, our statement is incorrect. It is possible to find
two distributions with the same variances and average degrees, which have
different values of $p_c$. One particular example is the following two
distributions $P_1(0)=0,P_1(1)=P_1(2)=P_1(3)=P_1(4)=P_1(5)=1/5$ and  $%
P_2(0)=P_2(3)=0,P_2(1)=P_2(5)=1/6,P_2(2)=P_2(4)=1/3$ which have $p_c$
respectively 0.683099 and 0.683657.

\section{General implications on the network robustness}

Finally, we will compare the robustness of CCN and RCN with the same degree
distributions. We will show that (\textit{i}) the value of $p_{c}$ for CCN
is always (except for RR networks) smaller than the $p_c$ for RCN and (%
\textit{ii}) for the same $p$, the value of the mutual giant component for
CCN is always (except for RR networks) larger than for RCN.

Eq. (\ref{tb}) for the randomly coupled networks can be rewritten as 
\begin{equation}
\frac{\lbrack 1-H(t)][1-G(t)]}{1-t}=\frac{1}{p},  \label{1pb}
\end{equation}%
The critical value of $p_{c}$ for randomly coupled networks can be
determined as the inverse maximal value of the left hand side of Eq. (\ref%
{1pb}). Our proposition (\textit{i}) is an obvious corollary of the
following proposition (\textit{iii}): for any $t\in \lbrack 0,1]$ the left
hand side of Eq. (\ref{1p}) is greater or equal than the left hand side of
Eq. (\ref{1pb}) (Fig.\ref{fig2}). Subtracting Eq. (\ref{1pb}) from Eq. (\ref%
{1p}) and applying relation (\ref{H}) between $G(t)$ and $H(t)$ we see that
the inequality stated in proposition (\textit{iii}) is equivalent to 
\begin{equation}
tG^{\prime}(t^2)-tG^{\prime }(t)+G(t)G^{\prime }(1)-G(t)G^{\prime }(t)\geq 0
\label{ineq}
\end{equation}%
We will prove Eq. (\ref{ineq}) using mathematical induction. We see that for
RR graphs for which the degree of every node is equal to $m$, i.e. $P(m)=1$,
Eq. (\ref{ineq}) is satisfied as an equality. Assuming that it is satisfied
for any degree distribution such that $P(k)=0$ for $k<m$ and $k>n\geq m$, we
will show that it is also satisfied for the degree distribution $\tilde{P}%
(k)=(1-b)P(k)$ for any $k$ except for $k=n+1$, for which $\tilde{P}(n+1)=b>0$%
. The generating function for this new distribution is obviously $~\tilde{G}%
=(1-b)G+bt^{n+1}$. After elementary algebra we can see that 
\begin{eqnarray}
& t \tilde{G}^{\prime}(t^2)-t\tilde{G}^{\prime}(t)+\tilde{G}(t)\tilde{G}^{\prime}(1)-\tilde{G}(t)\tilde{G^{\prime}}(t)=\\ &
(1-b)[t G^{\prime}(t^2)-tG^{\prime}(t)+G(t)G^{\prime}(1)-G(t)G^{\prime}(t)]+\\ &
b(1-b)(G(t)-t^{n+1})[(n+1-G^{\prime}(1))(1-t^n)+G^{\prime}(t)-G^{\prime}(1)t^{n}],
\end{eqnarray}
which proves Eq.~(\ref{ineq}) for $\tilde{G}$ provided it is true for $G$,
if we take into account the obvious inequalities $n+1>G^{\prime }(1)$, $%
1\geq t^{n}$, $G(t)\geq t^{n+1}$ and $G^{\prime }(t)\geq G^{\prime}(1)t^n$ for
any $t\in \lbrack 0,1]$. This concludes the proof of propositions (\textit{%
iii}) and (\textit{i}). Note that the equality sign in the above
inequalities and hence in inequality (\ref{ineq}) is realized only for $t=1$
and $t=0$ (if $P(0)=0$). Hence proposition (\textit{i}) always implies
rigorous inequality except for the case of RR graphs.

To prove the proposition (\textit{ii}) we first notice that the smallest
positive root of Eq.~(\ref{1p}), $t_{1}$, is always smaller than the
smallest positive root $t_{2}$ of Eq. (\ref{1pb}). This a is direct
consequence of proposition (\textit{iii}). Also we notice that the right
hand side of Eq. (\ref{mu}) is a monotonously decreasing function of $t$.
This can be shown by differentiation and comparing the terms of $G^{\prime }(t)$ and $tG^{\prime}
(t^2)$ corresponding to the same $P(k)$, namely $kP(k)t^{k-1}\geq kP(k)t^{2k-1}
$. Thus $\mu (t_{1})>\mu (t_{2})$. Finally, we state proposition (\textit{iv}%
): for the same value of $t$, the right hand side of Eq.~ (\ref{mu}), $\mu
(t)$, is greater or equal than the right hand side of Eq.~(\ref{mub}), $\mu
_{r}(t)$. One can prove this proposition using the same induction method  we
used to prove proposition (\textit{iii}). Combining these two results $\mu
(t_{1})>\mu (t_{2})\geq \mu _{r}(t_{2})$, which concludes the proof of
proposition (\textit{ii}).

Thus CCN are more robust than RCN with the same $P(k)$, but are still prone
to cascade failures and, then, to first order disintegration (only if $%
G^{\prime \prime }(1) <\infty$) as in case of randomly coupled networks.

\section{summary}

In this work we have studied the problem of failure of CCN, i.e. coupled
networks with coinciding degrees of mutually dependent nodes. We derive new
recursive equations [Eqs. (\ref{ti1}) and (\ref{Ni}) ] describing the
cascade of failures that are different from the analogous equations for RCN
studied in Ref.~\cite{Buldyrev}. We also find equations for the size of the
mutual giant component [Eqs.(\ref{mu}) and (\ref{t})], as well as the
efficient way of finding the critical fraction of nodes $p=p_c$ which must
survive the initial random failure for the mutual giant component not to
vanish, by finding the maximum of Eq. (\ref{1p}).

We show that if the second moment of the degree distribution is finite, CCN
disintegrate in a cascade of failures via a first order transition at which
the mutual giant component suddenly drops from a positive fraction above $%
p_c>0$ to zero below $p_c$. This behavior is analogous to the behavior of
RCN, with the only difference that RCN disintegrate via a first order 
transition even when the second moment of their degree distribution diverges.

Moreover, we show that CCN are always more robust than RCN with the same
degree distribution. In particular, we show that scale free CCN with $%
\lambda<3$ disintegrate via a second order phase transition in the same way
as non-interacting networks and thus are very resilient against random
failure. Namely, the mutual giant component for these networks exists at any 
$p>0$, but becomes infinitely small as $p\to 0$. Finally CCN become more
robust if their degree distribution becomes broader (provided the average
degree is constant). This behavior is the opposite of the behavior of RCN.

All our analytical predictions are confirmed by simulations of coupled
networks with large number of nodes ($N \geq 10^6$).

Based on our findings we conjecture that coupled networks with any
positively correlated degrees of mutually dependent nodes (and not just the
present case of fully coincidental degrees) are more robust that their
randomly coupled counterparts studied in Ref. \cite{Buldyrev} This can be
attributed to the fact that the correlation between the degrees of nodes
suppresses (or attenuates) the phenomenon of hubs becoming more vulnerable
by being dependent on low degree nodes in a coupled network.

\section{Acknowledgments}

We wish to thank DTRA for financial support and Dr. Robin Burk for
encouraging discussions. We acknowledge the partial support of this research
through the Dr. Bernard W. Gamson Computational Science Center at Yeshiva
College.

\end{document}